\documentclass[aps,prl,reprint,groupedaddress]{revtex4-1}
\usepackage{graphicx} 
\usepackage{dcolumn} 
\usepackage{bm}
\usepackage{hyperref}

\begin{document}

\title{Formation and dynamics of exotic hypernuclei in heavy-ion collisions}
\author{Zhao-Qing Feng}
\email{Corresponding author: fengzhq@scut.edu.cn}

\affiliation{School of Physics and Optoelectronics, South China University of Technology, Guangzhou 510640, China}

\date{\today}

\begin{abstract}
The dynamics of exotic hypernuclei in heavy-ion collisions has been investigated thoroughly with a microscopic transport model. All possible channels on hyperon ($\Lambda$, $\Sigma$ and $\Xi$) production near threshold energies are implemented in the transport model. The light complex fragments (Z$\leq$2) are constructed with the Wigner-function method. The classical phase-space coalescence is used for recognizing heavy nuclear and hyperfragments and the statistical model is taken for describing the decay process. The nuclear fragmentation reactions of the available experimental data from the ALADIN collaboration are well reproduced by the combined approach. It is found that the in-medium potentials of strange particles influence the strangeness production and fragment formation. The hyperfragments are mainly created in the projectile or target-like rapidity region and the yields are reduced about the 3-order magnitude in comparison to the nuclear fragments. The hypernuclear dynamics of HypHI data is well described with the model. The possible experiments for producing the neutron-rich hyperfragments at the high-intensity heavy-ion accelerator facility (HIAF) are discussed.

\begin{description}
\item[PACS number(s)]
21.80.+a, 24.10.Lx, 25.70.Pq     
\end{description}
\end{abstract}

\maketitle

\section{I. Introduction}

Heavy-ion collisions in terrestrial laboratory provide a unique possibility to create the strange particles in dense matter, which are available for investigating the in-medium properties of hadrons and nuclear equation of state (EOS) \cite{Ca99, Fu06, Li08, To10, Ha12}. Inclusion of the strangeness degree of freedom in nuclear medium extends the research activities in nuclear physics, in particular on the issues of the nuclear structure of hypernucleus and kaonic nucleus, hyperon-nucleon and hyperon-hyperon interactions, probing the in-medium properties of hadrons \cite{Gi95,Ha06,Ga16}. Moreover, hadrons with strangeness as essential ingredients influence the high-density nuclear EOS. The strangeness ingredient in dense matter softens the EoS at high-baryon densities, and consequently decreases the mass of neutron stars\cite{Ji13,We12}. Since the first observation of $\Lambda$-hypernuclide in nuclear multifragmentation reactions induced by cosmic rays in 1950s \cite{Da53}, a remarkable progress has been obtained in producing hypernuclides via different reaction mechanism, such as hadron (pion, K$^{\pm}$, proton, antiproton) induced reactions, bombarding the atomic nucleus with high-energy photons or electrons, and fragmentation reactions with high energy heavy-ion collisions. The proton-rich and neutron-rich hypernuclei in heavy-ion collisions is planned by the HypHI collaboration at the future facility for antiproton and ion research (FAIR) \cite{Ra16}. The strangeness physics is also attracted attention in the future experiments at the high-intensity heavy-ion accelerator facility (HIAF) \cite{Ya13,Ch19}.

The formation of hypernuclei in heavy-ion collisions is associated the creation of hyperons in hadron-hadron collisions, transportation in nuclear medium, fragment recognition and statistical decay. Up to now, several models have been established for describing the hypernucleus production, i.e., the statistical multifragmentation model (SMM) \cite{Bo07,Bo12}, statistical approach with a thermal source \cite{An11} and microscopic transport models with different recognition of hypernucleus \cite{Bo15,Le19}. Some interesting results are obtained for describing the hypernucleus formation, i.e., the hyperfragment yields, excitation function of hyperfragment production, multiple strangeness hypernucleus etc. More sophisticated investigation of hypernuclear dynamics is needed in theories.

In this work, the Lanzhou quantum molecular dynamics (LQMD) transport model is extended for investigating the strangeness production and hypernuclear dynamics in heavy-ion collisions near threshold energies. The article is organized as follows. In section II we give a brief description of the model for the strangeness production and fragment recognition. The calculated results and discussion are presented in section III. Summary and perspective are outlined in section IV.

\section{II. the transport model and fragment recognition}

In the LQMD transport model, the production of resonances with the mass below 2 GeV, hyperons ($\Lambda$, $\Sigma$, $\Xi$) and mesons ($\pi$, $\eta$, $K$, $\overline{K}$, $\rho$, $\omega$) is coupled in the reaction channels via meson-baryon and baryon-baryon collisions \cite{Fe11,Fe18}. The temporal evolutions of nucleons are described by Hamilton's equations of motion under the self-consistently generated two-body and three-body interaction potential with the Skyrme force. The chiral effective Lagrangian and relativistic mean-field theories are applied for evaluating the one-body potentials of mesons and hyperons in nuclear medium. The hyperon mean-field potential is constructed on the basis of the light-quark counting rule. The self-energies of hyperons are assumed to be two thirds of that experienced by nucleons. Thus, the in-medium dispersion relation reads
\begin{equation}
\omega(\textbf{p}_{i},\rho_{i})=\sqrt{(m_{H}+\Sigma_{S}^{H})^{2}+\textbf{p}_{i}^{2}} + \Sigma_{V}^{H}
\end{equation}
with $\Sigma_{S}^{H}= 2 \Sigma_{S}^{N}/3$ and $\Sigma_{V}^{H}= 2 \Sigma_{V}^{N}/3$, which leads to the attractive interaction being the values of -32 MeV and -16 MeV for $\Lambda$ and $\Xi$ at the normal nuclear density, respectively. The nuclear scalar $\Sigma_{S}^{N}$ and vector $\Sigma_{V}^{N}$ self-energies are computed from the well-known relativistic mean-field model with the NL3 parameter \cite{La97}.

Production and decay of the resonances below the mass of 2 GeV have been included in the model \cite{Fe18}. All possible channels of $\Xi$ production in hadron-hadron collisions are implemented in this work. The strange particles are created in the direct process by the channels as follows
\begin{eqnarray}
&& BB \rightarrow BYK,  BB \rightarrow BBK\overline{K},  B\pi(\eta) \rightarrow YK,  YK \rightarrow B\pi,     \nonumber \\
&& B\pi \rightarrow NK\overline{K}, Y\pi \rightarrow B\overline{K}, \quad  B\overline{K} \rightarrow Y\pi, \quad YN \rightarrow \overline{K}NN,  \nonumber \\
&& BB \rightarrow B\Xi KK, \overline{K}B \leftrightarrow K\Xi, YY \leftrightarrow N\Xi, \overline{K}Y \leftrightarrow \pi\Xi
\end{eqnarray}
Here the symbols corresponding to B(N, $\triangle$, N$^{\ast}$), Y($\Lambda$, $\Sigma$), $\Xi(\Xi^{0}, \Xi^{-}$), $\pi(\pi^{-}, \pi^{0}, \pi^{+})$, K(K$^{0}$, K$^{+}$), $\overline{K}$($\overline{K}^{0}$, K$^{-}$). The elementary cross sections are parameterized by fitting the available experimental data and the Clebsch-Gordan coefficients for the isospin channels. Furthermore, the elastic scattering and strangeness-exchange reaction between strangeness and baryons have been considered through the channels of $KB \rightarrow KB$, $YB \rightarrow YB$ and $\overline{K}B \rightarrow \overline{K}B$ and we use the parametrizations in Ref. \cite{Cu90}. The charge-exchange reactions between the $KN \rightarrow KN$ and $YN \rightarrow YN$ channels are included by using the same cross sections with the elastic scattering, such as $K^{0}p\rightarrow K^{+}n$, $K^{+}n\rightarrow K^{0}p$ etc \cite{Fe13}.

The primary fragments with $Z\geq$3 are recognized in phase space with a coalescence model, in which the nucleons at freeze-out stage are considered to belong to one cluster with the relative momentum smaller than $P_{0}$ and with the relative distance smaller than $R_{0}$ (here $P_{0}$ = 200 MeV/c and $R_{0}$ = 3 fm). A larger relative distance ($R_{0}$ = 5 fm) is taken into account for constructing the hyperfragments, which is caused from the fact that the weakly bound of hypernucleus with a bigger rms (root-mean-square) radius, e.g., 5 fm rms for $^{3}_{\Lambda}$H and 1.74 fm for $^{3}$He \cite{Ar04}. Actually, the influence of the coalescence parameters on the final fragments is small because the lager coalescence distance increases the excitation energy of primary fragment and enable more probability in the de-excitation process. The excitation energy is evaluated as the difference of binding energies between the excited fragment and the Bethe-Weiz\"{a}cker mass formula for nuclear fragments. The generalized mass formula with SU(6) symmetry breaking is used for calculating the hypernuclear binding energy \cite{Sa06}. The de-excitation process of the nuclear fragment and hyperfragments is described by the GEMINI code \cite{Ch88}, in which the channels of $\gamma$, light complex clusters (n, p, $\alpha$ etc) and binary fragments are selected by the Monte Carlo procedure via the decay width. The decay widths of light particles with Z$\leq$2 and the binary decay are calculated by the Hauser-Feshbach formalism \cite{Ha52} and transition state formalism \cite{Mo75}, respectively. The hyperon decay width is also evaluated by the Hauser-Feshbach approach the phenomenological hyperon binding energy \cite{Fe20}. In the LQMD model, the binding energy of the primary fragment is calculated by the internal motion energy and interaction potential as
\begin{eqnarray}
E_{B}(Z_{i},N_{i})= &&\sum_{j}\sqrt{p_{j}^{2}+m_{j}^{2}}-m_{j}          \nonumber \\
&& +\frac{1}{2}\sum_{j,k,k\neq j}\int f_{j}(\textbf{r},\textbf{p},t)f_{k}(\textbf{r}^{\prime},\textbf{p}^{\prime},t)  \nonumber \\
&& \times v(\textbf{r},\textbf{r}^{\prime},\textbf{p},\textbf{p}^{\prime})    d\textbf{r}d\textbf{r}^{\prime} d\textbf{p}d\textbf{p}^{\prime}               \nonumber \\
&& +\frac{1}{6}\sum_{j,k,l}\sum_{k\neq j, k\neq l,j\neq l}
\int f_{j}(\textbf{r},\textbf{p},t) f_{k}(\textbf{r}^{\prime},\textbf{p}^{\prime},t)                            \nonumber \\
&& \times  f_{l}(\textbf{r}^{\prime\prime},\textbf{p}^{\prime\prime},t) v(\textbf{r},\textbf{r}^{\prime},\textbf{r}^{\prime\prime},
\textbf{p},\textbf{p}^{\prime},\textbf{p}^{\prime\prime})             \nonumber \\
&& \times  d\textbf{r}d\textbf{r}^{\prime}d\textbf{r}^{\prime\prime} d\textbf{p}d\textbf{p}^{\prime}d\textbf{p}^{\prime\prime},
\end{eqnarray}
where the $\textbf{r}, \textbf{p}$ are the nucleon position in the center of mass of the $i-$th fragment $(Z_{i}, N_{i})$. We count the binding energy of hyperfragment with $E_{B}(Z_{i},N_{i},N_{Y})=E_{B}(Z_{i},N_{i})+\sum_{j=1}^{N_{Y}}\omega(\textbf{p}_{j},\rho_{j})-m_{H}$ with the $Z_{i}$, $N_{i}$ and $N_{Y}$ being the proton, neutron and hyperon numbers, respectively.

For the fragment with $Z\leq$2, the Wigner phase-space density at freeze out is used to evaluate the probability of fragment formation. It is assumed that the cold clusters are formed at freeze out. The momentum distribution of a cluster with $M$ nucleons and $Z$ protons for a system with $A$ nucleons is given by
\begin{eqnarray}
\frac{dN_{M}}{d^{3}P}=&& G_{M}{A \choose M} {M \choose Z}\frac{1}{A^{M}}\int \prod_{i=1}^{Z}f_{p}(\textbf{r}_{i},\textbf{p}_{i}) \prod_{i=Z+1}^{M}f_{n}(\textbf{r}_{i},\textbf{p}_{i})                  \nonumber \\
&& \times  \rho^{W}(\textbf{r}_{k_{1}},\textbf{p}_{k_{1}},...,\textbf{r}_{k_{M-1}},\textbf{p}_{k_{M-1}})             \nonumber \\
&& \times  \delta(\textbf{P}-(\textbf{p}_{1}+...+\textbf{p}_{M}))d\textbf{r}_{1}d\textbf{p}_{1}...d\textbf{r}_{M}d\textbf{p}_{M}.
\end{eqnarray}
Here the $f_{n}$ and $f_{p}$ are the neutron and proton phase-space density, which are obtained by performing Wigner transformation based on Gaussian wave packet. The relative coordinate $\textbf{r}_{k_{1}}, ..., \textbf{r}_{k_{M-1}}$ and momentum $\textbf{p}_{k_{1}}, ...,\textbf{p}_{k_{M-1}}$ in the $M-$nucleon rest frame are used for calculating the Wigner density $\rho^{W}$ \cite{Ma97,Ch03}. The spin-isospin statistical factor $G_{M}$ is 3/8, 1/12 and 1/96 corresponding to M=2, 3 and 4, respectively. The root-mean-square radii of intending a cluster is needed for the Wigner density, i.e., 1.61 fm and 1.74 fm for triton and $^{3}$He, loosely bound for hypernuclide.

\section{III. Results and discussion}

Particles production in heavy-ion collisions provides the possibility for extracting the high-density nuclear matter properties and investigating the behaviour of in-medium hadrons. Strangeness physics with heavy-ion collisions manifests some interesting phenomena, such as the bound state of multiple baryons, multistrangeness hypernucleus, interaction between strange particles etc. To explore the emission mechanism of strange particles, we calculated the rapidity and transverse momentum spectra in nuclear collisions. Shown in Fig. 1 is the strangeness production in the reaction of $^{124}$Sn+$^{124}$Sn at the incident energy of 2 GeV/nucleon. The interaction potential of strange particle and nucleon influences the strangeness production and dynamics, i.e., the 30 $\%$ reduction for K$^{+}$ yields with the potential. In the mid-rapidity region, the strangeness production is reduced by the optical potential and escaped from the nuclear medium at the early stage in heavy-ion collisions, in which the high-density properties of hadronic matter can be extracted from the particle spectra.
The transverse momentum spectra of strange particles are calculated as shown in Fig. 2. It is obvious that the inclusion of the optical potential leads to the reduction of low-momentum particles. The low-momentum hyperons can be easily captured by surrounding nucleons to form hyperfragments. On the other hand, the high-momentum strange particles are nice probes for extracting the high-density matter properties, i.e., symmetry energy, effective mass, short-range range interaction etc.

\begin{figure*}
\includegraphics[width=16 cm]{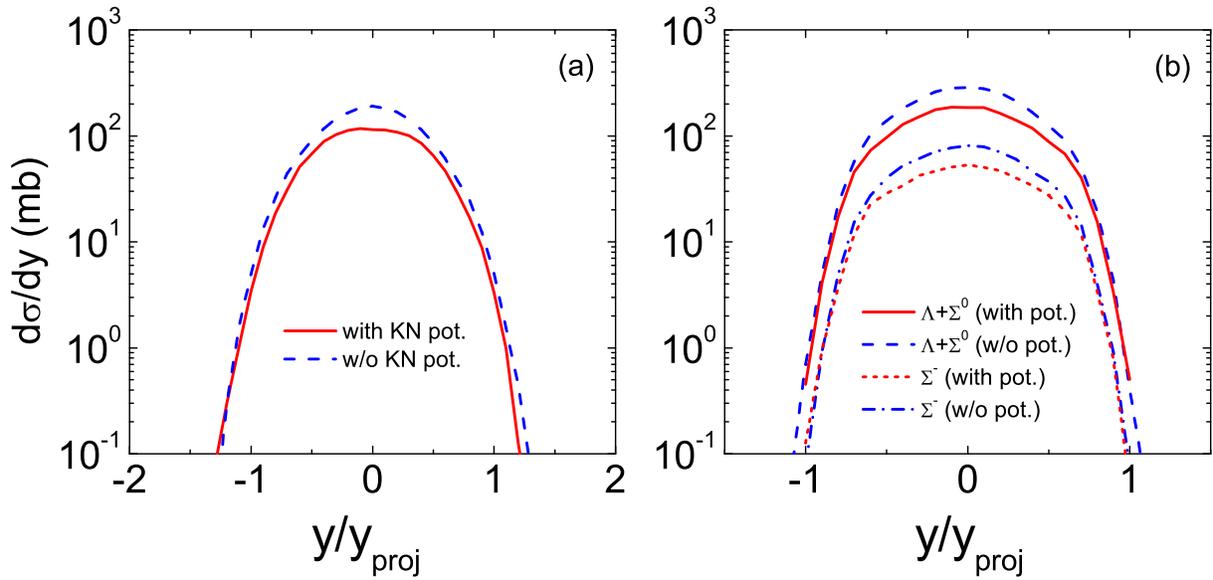}
\caption{Rapidity distribution of K$^{+}$ (a) and hyperons (b) in collisions of $^{124}$Sn+$^{124}$Sn at the incident energy of 2 GeV/nucleon.}
\end{figure*}

\begin{figure*}
\includegraphics[width=16 cm]{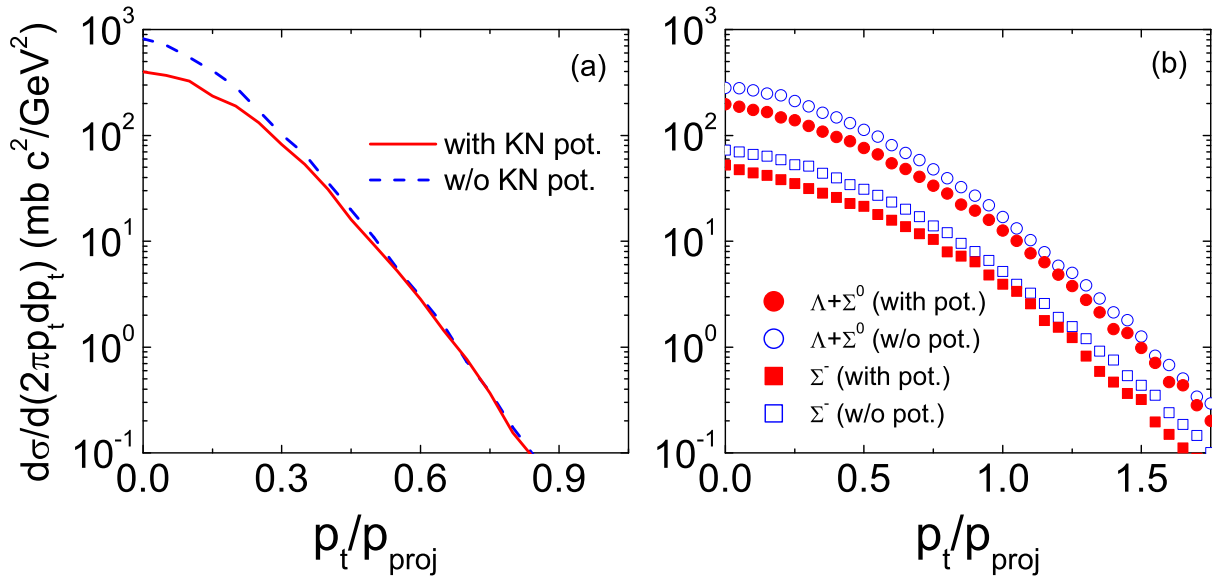}
\caption{Influence of the in-medium potentials on the production of kaons and hyperons in the reaction of $^{124}$Sn+$^{124}$Sn.}
\end{figure*}

The nuclear fragmentation reactions have been extensively investigated both in experiments and in theories, in particular on the issues of spinodal multifragmentation, liquid-gas phase transition, properties of highly excited nuclei, symmetry energy at subsaturation densities etc \cite{Ch04,Co93,Po95,Wu98,Ma99}. The intermediate mass fragments (IMFs) with $3\leq Z \leq 30$ produced in Fermi-energy heavy-ion collisions has been investigated for extracting the nuclear equation of state, liquid-gas phase transition and density dependence of symmetry energy \cite{Li04,Fe16}, where the composite system is formed at excitation energies of 10-20 MeV per nucleon. The fluctuation and explosive decay of the excited system dominate the IMF production. The quantity Z$_{bound}$ is defined as the sum of fragments with $Z_{i}\geq$2. As a test of the combined approach, shown in Fig. 3 is the averaged IMFs correlated with the Z$_{bound}$ and impact parameter $b$ in the reaction $^{197}$Au+$^{197}$Au at the incident energy of 1 GeV/nucleon and compared with the data from the ALADIN spectrometer \cite{Sc96}. The reaction system is evolved until to 300 fm/c with the momentum-dependent interaction in the LQMD model, in which the nucleon-nucleon scattering is finished and the chemical equilibrium is reached (freeze out). The dominant decay channels are $\alpha$, proton and neutron.

\begin{figure*}
\includegraphics[width=16 cm]{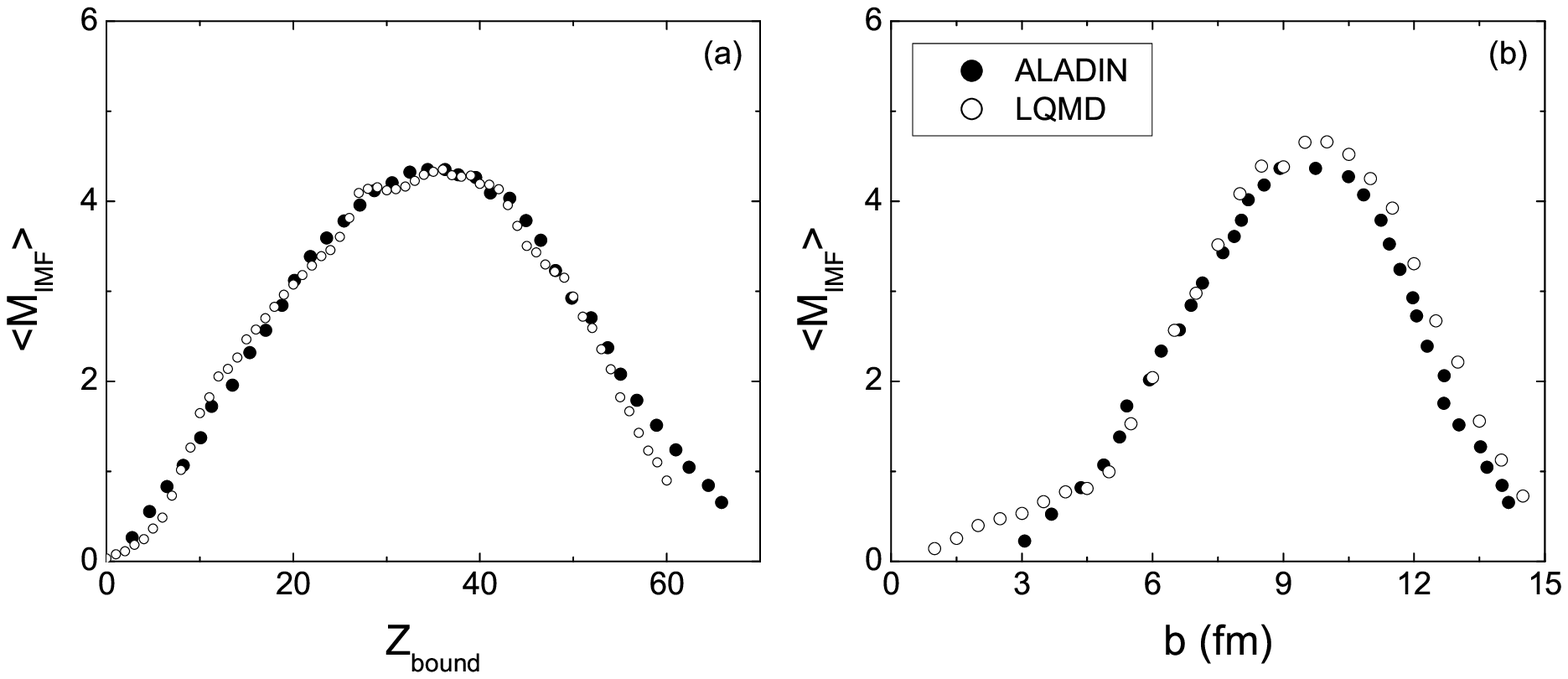}
\caption{Mean multiplicity of intermediate mass fragments $<M_{IMF}>$ as functions of the Z$_{bound}$ and impact parameter $b$ in the reaction $^{197}$Au+$^{197}$Au at the incident energy of 1 GeV/nucleon and compared with the data from the ALADIN spectrometer \cite{Sc96}.}
\end{figure*}

\begin{figure*}
\includegraphics[width=16 cm]{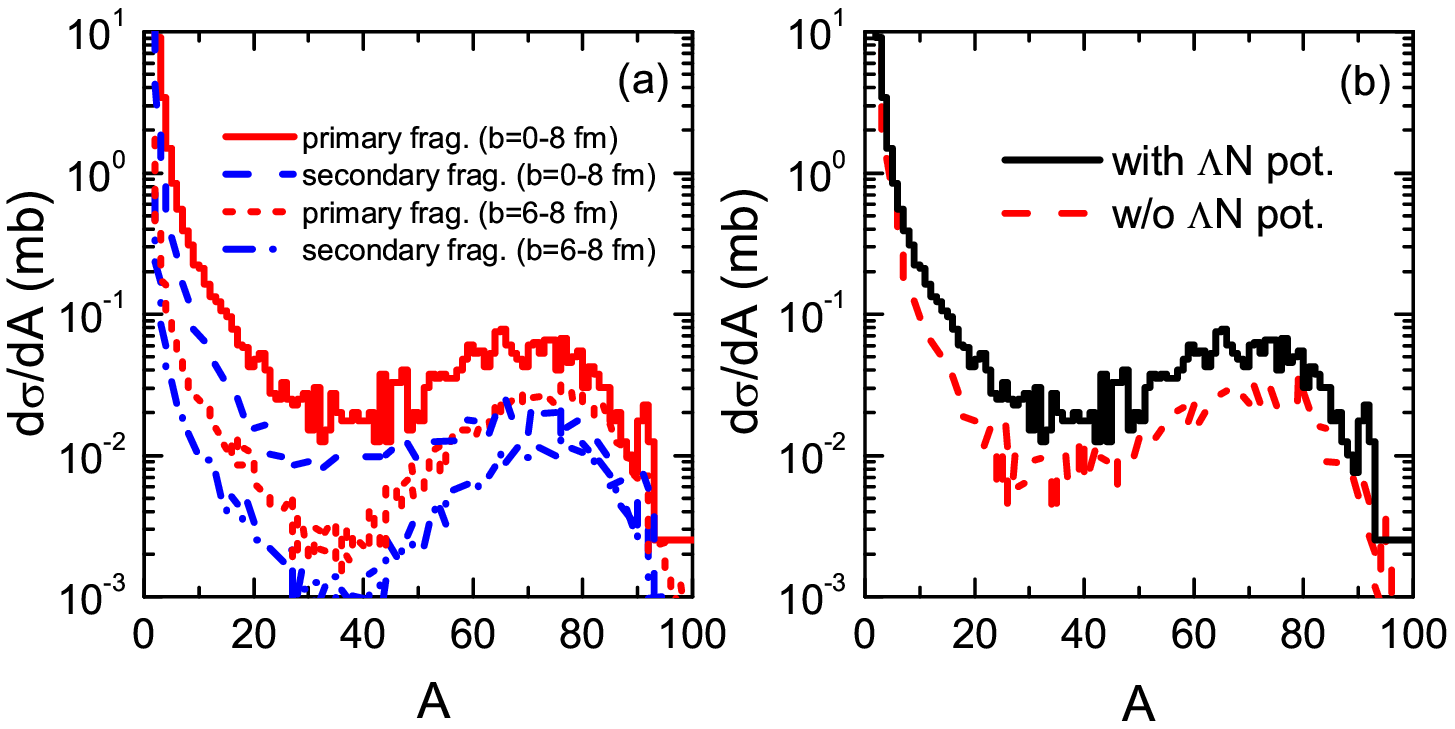}
\caption{Influence of the statistical decay and the $\Lambda$-nucleon potential on the hyperfragment production in collisions of $^{124}$Sn+$^{124}$Sn at 2 GeV/nucleon.}
\end{figure*}

The hyperfragments constructed in phase space are excited, in which the hyperon-nucleon interaction potential influences the dynamics of hyperon transportation in nuclear medium and consequently the bound fragments by capturing a hyperon. The GEMINI code is modified for modeling the decay of excited hypernucleus. Shown in Fig. 4 is a comparison of the primary and secondary fragments produced in the reaction of $^{124}$Sn+$^{124}$Sn at the incident energy of 2 GeV/nucleon. It is obvious that the peripheral collisions mainly contribute the large mass hyperfragments. The yields of hyperfragments are reduced by implementing the statistical decay. The attractive $\Lambda$-nucleon potential is available for enhancing the hyperfragment production. It should be noticed that the classical method is used for constructing the primary hyperfragments in phase space in this work. A quantal approach for recognizing the fragments is still expected, in which the root-mean-square radii and binding energy should be taken into account by relying on the nuclear structure calculations. The Minimum Spanning Tree (MST) and the Simulated Annealing Cluster Algorithm (SACA) algorithms were proposed for constructing nuclear clusters and hypernuclei within the transport model PHQMD (Parton Hadron Quantum Molecular Dynamics) \cite{Le19, Ai19}. The nuclear dynamics of cluster and hypernucleus production in fragmentation reactions were investigated thoroughly.

The nuclear chart is extended to the three dimensional structure by implementing strangeness degree of freedom, which exhibits some interesting phenomena, i.e., the neutral bound state (nn$\Lambda$, n$\Lambda\Lambda$) \cite{Ne05,Ga13}, new spectroscopy etc. The heavy-ion collisions provide a unique way to produce the neutron-rich or proton-rich hypernuclides in the terrestrial laboratories. Shown in Fig. 5 is the light hypernuclides $^{3}_{\Lambda}H$ and $^{4}_{\Lambda}H$ produced in the reaction $^{6}$Li+$^{12}$C at an incident energy of 2 GeV/nucleon and compared with the experimental data by the HypHI collaboration in the projectile spectator region \cite{Ra15}. The Wigner density approach is used for recognizing the light hypernuclei at freeze out. It is noticed that the hyperfragments are likely to be formed in the peripheral collisions within the low-momentum region. Very narrow rapidity spectra are found, which are caused from that the hyperon is captured by spectator nucleons. To create the neutron-rich hypernucleus, the neutron diffusion to the spectator region is expected. The HypHI data were also analyzed by two approaches of the QMD model with FRIGA (Fragment Recognition In General Application) \cite{Le19} and the dubna cascade model followed by a statistical code \cite{Su18}. In this work, the preequilibrium emissions of light clusters and hypernuclei from 3-body or 4-body collisions are not included. Further modifications on light hypernuclear production in transport model are in progress.

\begin{figure*}
\includegraphics[width=16 cm]{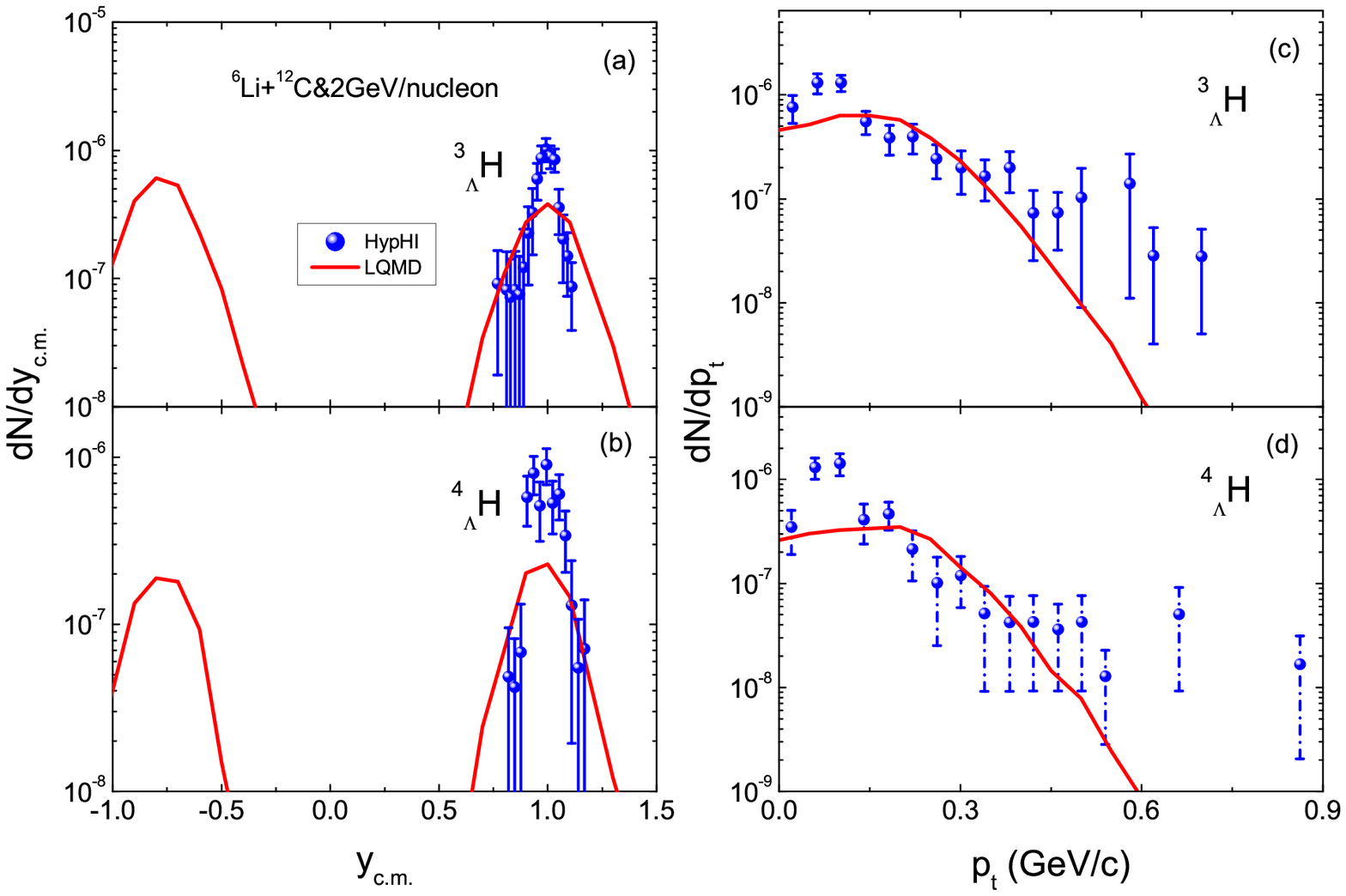}
\caption{Rapidity and transverse momentum spectra of $^{3}_{\Lambda}H$ and $^{4}_{\Lambda}H$ produced in the reaction $^{6}$Li+$^{12}$C at an incident energy of 2 GeV/nucleon and compared with the HypHI data \cite{Ra15}.}
\end{figure*}

Formation mechanism of fragments with strangeness in high-energy heavy-ion collisions has been investigated within the LQMD transport approach combined with the statistical model (GEMINI). The production and dynamics of hyperons is described within the LQMD model. The classical coalescence model is used for constructing the primary fragments and the hyperon capture by residual nucleons. The combined approach is used to describe the formation of hyperfragments with Z$\geq$3. The production of hypernuclei is associated to the hyperon production, hyperon-nucleon and hyperon-hyperon interactions, capture of hyperons by nucleonic fragments, and decay of excited hyper-fragments. The investigation of hypernucleus properties is also an essential way for extracting the in-medium information of hyperons. To form the hyperfragments in heavy-ion collisions, the incident energy is chosen to be enough for creating hyperons, but not too high so that the hyperon evolution is enough for capturing by surrounding nucleons. At the near threshold energies, the reaction channels of $BB \rightarrow BYK$ and $\pi(\eta)B \rightarrow KY$ dominate the hyperon production. Usually, the hyperons are created in the domain of the dense nuclear medium. The hyperon-nucleon potential impacts the hyperon dynamics and hyperfragment formation. Calculations from a statistical model gave that the beam energy regime of 3-5 GeV/nucleon is available for producing hypernuclei \cite{An11}. The production and structure studies of neutron-rich and even double-strangeness hypernuclides have been planned at the HIAF in China. Shown in Fig. 6 is the nuclear fragments and $\Lambda-$hyperfragments produced in collisions of $^{112}$Sn+$^{112}$Sn, $^{124}$Sn +$^{124}$Sn and $^{132}$Sn +$^{124}$Sn at the incident energy of 2 GeV/nucleon and within the collision centrality of 0-8 fm. The mass and charge distribution trends of nuclear fragments and hyperfragments are very similar. The yields almost sustain constant in the mass region A=20$\sim$70 (or Z=10$\sim$30) because of the contribution of peripheral collisions. The production of hyperfragments is reduced the 2-3 order magnitude to the nuclear fragments in the large mass domain. The light hyperfragments are feasible via heavy-ion collisions. The dynamics of fragments is calculated as shown in Fig. 7. A broad distribution is obvious for the nuclear fragments. Symmetric structure appears for both the nuclear fragments and hyperfragments. The fragments tend to be formed in the projectile or target-like region with increasing the charge number.

\begin{figure*}
\includegraphics[width=16 cm]{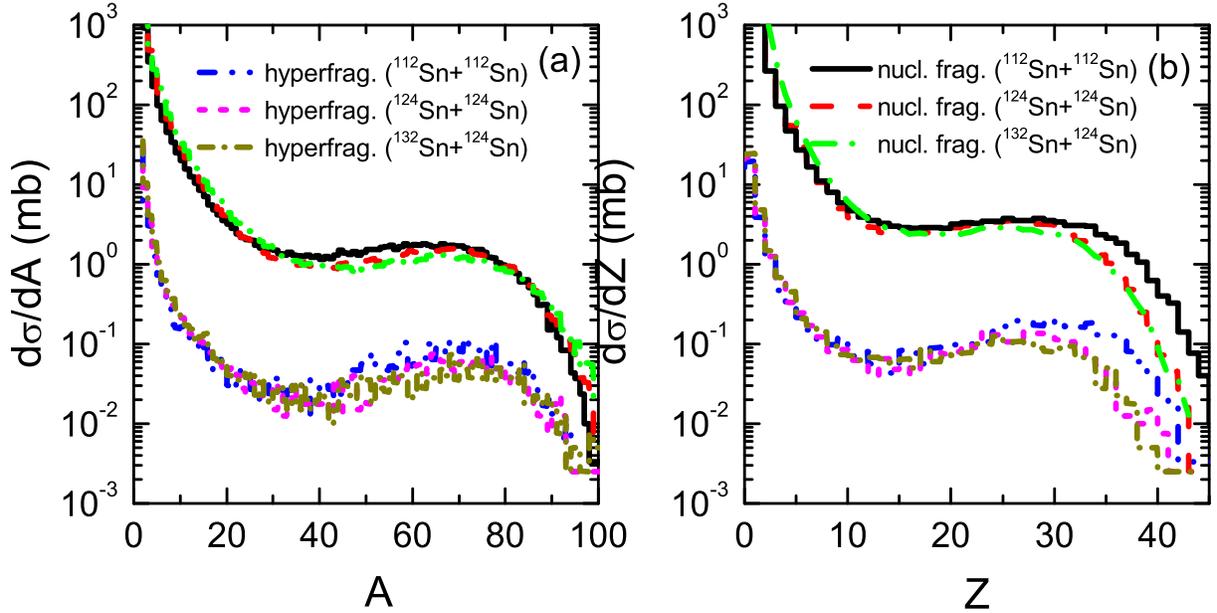}
\caption{Comparison of the nuclear fragments and $\Lambda-$hyperfragments as functions of mass and charged numbers in collisions of $^{112}$Sn+$^{112}$Sn, $^{124}$Sn+$^{124}$Sn and $^{132}$Sn+$^{124}$Sn at 2$\emph{A}$ GeV, respectively.}
\end{figure*}

\begin{figure*}
\includegraphics[width=16 cm]{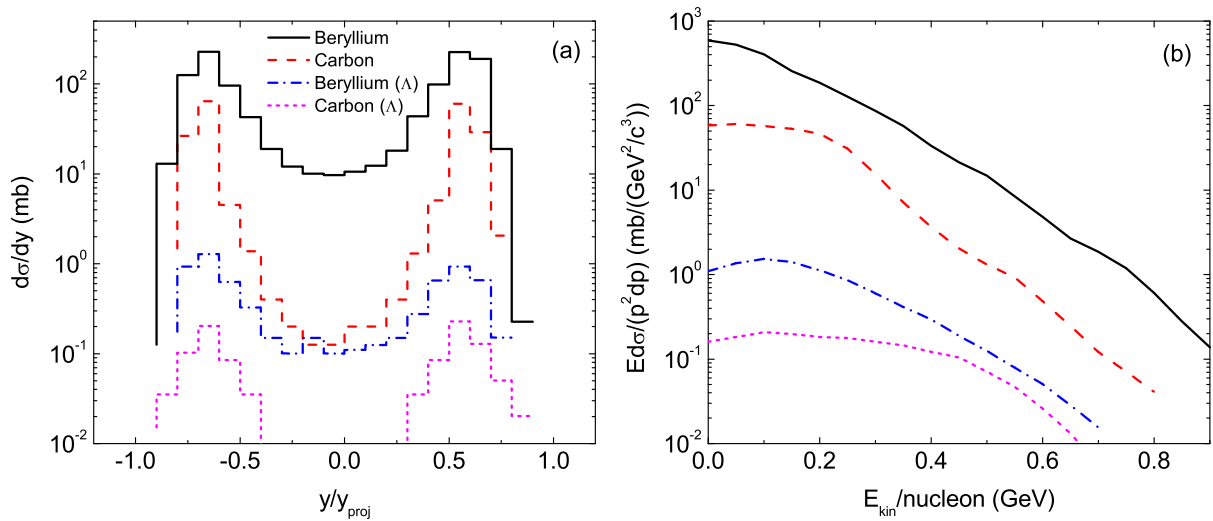}
\caption{Rapidity and kinetic energy spectra of nucleonic and hypernuclear fragments for Beryllium and Carbon isotope production in the $^{124}$Sn+$^{124}$Sn reaction at 2$\emph{A}$ GeV.}
\end{figure*}

The hypernuclear physics is one of the topical issues in the future HIAF facility, in particular the exotic hypernuclei in extremely neutron-rich or proton-rich region, multiple strangeness nuclei, high-density symmetry energy from the isospin ratios of strange particles etc. The first experiment is planned for investigating the light hypernuclei in collisions of $^{20}$Ne+$^{12}$C at the beam energy of 4.25 GeV/nucleon. We analyzed the dynamics of light hypernuclei. Shown in Fig. 8 is a comparison of light nuclei, hypernuclei and free $\Lambda$ produced in collisions of $^{20}$Ne+$^{12}$C. It is pronounced that the clusters are formed in the projectile and target-like rapidity region. The free $\Lambda$ escape the 'fire ball' formed in nuclear collisions and the dominant emission in the midrapidity region. The hypernuclei are produced within the narrow rapidity and low kinetic energies. The kinematics of hyperfragments is helpful in managing the detector system in the future experiments.

\begin{figure*}
\includegraphics[width=16 cm]{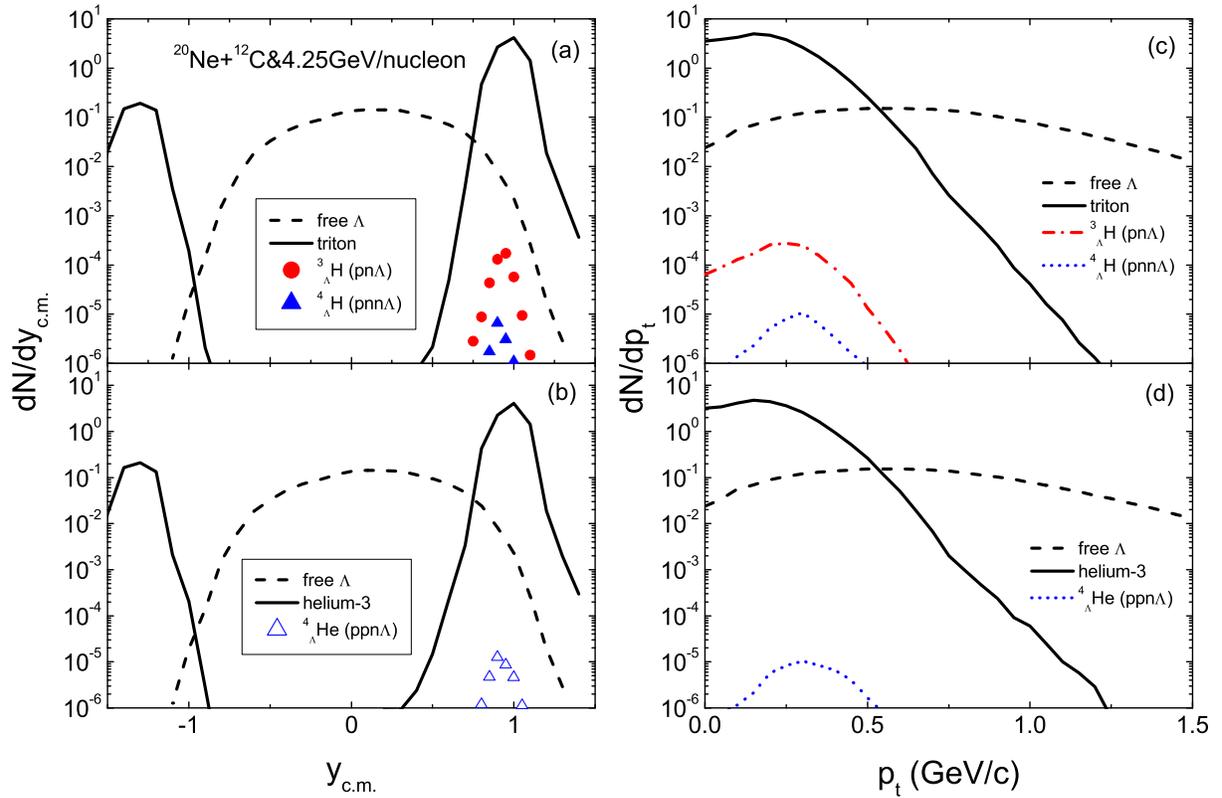}
\caption{Comparison of rapidity and transverse momentum distributions for $^{3}$H, $^{3}$He, $^{3}_{\Lambda}$H, $^{4}_{\Lambda}$H, $^{4}_{\Lambda}$He and free $\Lambda$ produced in collisions of $^{20}$Ne+$^{12}$C at the incident energy of 4.25 GeV/nucleon.}
\end{figure*}

\section{IV. Conclusions}

In summary, the formation mechanism of hypernuclei in heavy-ion collisions is investigated within the LQMD transport model combined the coalescence approach and statistical model. The dynamics of strange particles and hypernucleus formation is influenced by the optical potentials in nuclear medium. The combined approach can nicely describe the fragmentation reactions of the available experimental data from the ALADIN collaboration and the hypernucleus data from the HypHI collaboration. The strange particles are mainly emitted in the mid-rapidity region and with the broad momentum range. However, the hyperfragments are created in the projectile or target-like rapidity region by capturing of the spectator nucleons and the yields are reduced about the 3-order magnitude in comparison to the nuclear fragments. The possible experiments for producing the neutron-rich hyperfragments at HIAF are discussed.

\section{Acknowledgements}

This work was supported by the National Natural Science Foundation of China (Projects No. 11722546 and No. 11675226) and the Talent Program of South China University of Technology.


\begin{thebibliography}{99}

\bibitem{Ca99} W. Cassing and E. L. Bratkovskaya, Phys. Rep. \textbf{308}, 65 (1999).
\bibitem{Fu06} C. Fuchs, Prog. Part. Nucl. Phys. \textbf{56}, 1 (2006).
\bibitem{Li08} B. A. Li, L. W. Chen, and C. M. Ko, Phys. Rep. \textbf{464}, 113 (2008).
\bibitem{To10} M. Di Toro, V Baran, M Colonna, and V Greco, J. Phys. G: Nucl. Part. Phys. \textbf{37}, 083101 (2010).
\bibitem{Ha12} C. Hartnack, H. Oeschler, Y. Leifels, E. Bratkovskaya, and J. Aichelin, Phys. Rep. \textbf{510}, 119 (2012).
\bibitem{Gi95} B.E. Gibson and E.V. Hungerford III, Phys. Rep. \textbf{257}, 349 (1995).
\bibitem{Ha06} O. Hashimoto and H. Tamura, Prog. Part. Nucl. Phys. \textbf{57}, 564 (2006).
\bibitem{Ga16} A. Gal, E. V. Hungerford, and D. J. Millener, Rev. Mod. Phys. \textbf{88}, 035004 (2016).
\bibitem{Ji13} W. Z. Jiang, R. Y. Yang, and D. R. Zhang, Phys. Rev. C \textbf{87}, 064314 (2013).
\bibitem{We12} S. Weissenborn, D. Chatterjee, and J. Schaffner-Bielich, Nucl. Phys. A \textbf{881}, 62 (2012).
\bibitem{Da53} M. Danysz and J. Pniewski, Philos. Mag. \textbf{44}, 348 (1953).
\bibitem{Ra16} C. Rappold and J. L\'{o}pez-Fidalgo, Phys. Rev. C \textbf{94}, 044616 (2016).
\bibitem{Ya13} J. C. Yang, J. W. Xia, G. Q. Xiao \emph{et al.}, Nucl. Instrum. Methods B \textbf{317}, 263 (2013).
\bibitem{Ch19} X. Chen, J. C. Yang, J. W. Xia \emph{et al.}, Nucl. Instrum. Methods A \textbf{920}, 37 (2019).
\bibitem{Bo07} A. S. Botvina and J. Pochodzalla, Phys. Rev. C \textbf{76}, 024909 (2007).
\bibitem{Bo12} A. S. Botvina, K. K. Gudima, J. Steinheimer \emph{et al.}, Nucl. Phys. A \textbf{881}, 228 (2012).
\bibitem{An11} A. Andronic, P. Braun-Munzinger, J. Stachel, H. St\"{o}cker, Phys. Lett. B \textbf{697}, 203 (2011).
\bibitem{Bo15} A. S. Botvina, J. Steinheimer, E. Bratkovskaya, M. Bleicher, J. Pochodzalla, Phys. Lett. B \textbf{742}, 7 (2015).
\bibitem{Le19} A. Le F\`{e}vre, J. Aichelin, C. Hartnack, and Y. Leifels, Phys. Rev. C \textbf{100}, 034904 (2019).
\bibitem{Fe11} Z. Q. Feng, Phys. Rev. C \textbf{84}, 024610 (2011).
\bibitem{Fe18} Z. Q. Feng, Nucl. Sci. Tech., \textbf{29}, 40 (2018).
\bibitem{La97} G. A. Lalazissis, J. K\"{o}nig, and P. Ring, Phys. Rev. C \textbf{55}, 540 (1997).
\bibitem{Cu90} J. Cugnon, P. Deneye, and J. Vandermeulen, Phys. Rev. C \textbf{41}, 1701 (1990).
\bibitem{Fe13} Z. Q. Feng, Nucl. Phys. A \textbf{919}, 32 (2013).
\bibitem{Ar04}  T. A. Armstrong \emph{et al.}, Phys. Rev. C \textbf{70}, 024902 (2004).
\bibitem{Sa06} C. Samanta, P. Roy Chowdhury, D. N. Basu, J. Phys. G: Nucl. Part. Phys. \textbf{32}, 363 (2006); C. Samanta, J. Phys. G: Nucl. Part. Phys. \textbf{37}, 075104 (2010).
\bibitem{Ch88} R. J. Charity \emph{et al.}, Nucl. Phys. A \textbf{483}, 371 (1988).
\bibitem{Ha52} H. Hauser and H. Feshbach, Phys. Rev. \textbf{87}, 366 (1952).
\bibitem{Mo75} L. G. Moretto, Nucl. Phys. A \textbf{247}, 211 (1975).
\bibitem{Fe20} Z. Q. Feng, Phys. Rev. C \textbf{101}, 014605 (2020).
\bibitem{Ma97} R. Mattiello, H. Sorge, H. St\"{o}cker, and W. Greiner, Phys. Rev. C \textbf{55}, 1443 (1997).
\bibitem{Ch03} L. W. Chen, C. M. Ko, B. A. Li, Nucl. Phys. A \textbf{729}, 809 (2003).
\bibitem{Ch04} P. Chomaz, M. Colonna, and J. Randrup, Phys. Rep. \textbf{389}, 263 (2004).
\bibitem{Co93} M. Colonna, M. Di Toro, A. Guarnera, V. Latora, and A. Smerzi, Phys. Lett. B \textbf{307}, 273 (1993); M. Colonna, M. Di Toro, and A. Guarnera, Nucl. Phys. A \textbf{580}, 312 (1994).
\bibitem{Po95} J. Pochatlzalla \emph{et al.}, Phys. Rev. Lett. \textbf{75}, 1040 (1995).
\bibitem{Wu98} H.-Y. Wu, G.-X. Dai, G.-M. Jin \emph{et al.}, Phys. Rev. C \textbf{57}, 3178 (1998); H.-Y. Wu \emph{et al.}, Phys. Lett. B \textbf{538}, 39 (2002).
\bibitem{Ma99} Y. G. Ma, Phys. Rev. Lett. \textbf{83}, 3617 (1999).
\bibitem{Li04} T. X. Liu, M. J. van Goethem, X. D. Liu \emph{et al.}, Phys. Rev. C \textbf{69}, 014603 (2004).
\bibitem{Fe16} Z. Q. Feng, Phys. Rev. C \textbf{94}, 014609 (2016).
\bibitem{Sc96} A. Sch\"{u}ttauf, W. D. Kunze, A. W\"{o}rner \emph{et al.}, Nucl. Phys. A \textbf{607}, 457 (1996).
\bibitem{Ai19} J. Aichelin, E. Bratkovskaya, A. Le F\`{e}vre, V. Kireyeu, V. Kolesnikov, Y. Leifels, V. Voronyuk, and G. Coci, arXiv:
1907.03860.
\bibitem{Ne05} H. Nemura, S. Shinmura, Y. Akaishi, and Khin Swe Myint, Phys. Rev. Lett. \textbf{94}, 202502 (2005).
\bibitem{Ga13} H. Garcilazo and A. Valcarce, Phys. Rev. Lett. \textbf{110}, 012503 (2013).
\bibitem{Ra15} C. Rappold et al., Phys. Lett. B \textbf{747}, 129 (2015).
\bibitem{Su18} Y. L. Sun, A. S. Botvina, A. Obertelli, A. Corsi, and M. Bleicher, Phys. Rev. C \textbf{98}, 024903 (2018).

\end{thebibliography}
\end{document}